# Synchronization of Kuramoto oscillators on Knots

Amelia Carolina Sparavigna[1]

[1] Department of Applied Science and Technology, Politecnico di Torino, Corso Duca degli Abruzzi 24, Torino, Italy

**Abstract**: A knot is a circle embedded in the space. Projecting a knot on a plane, we obtain a diagram which is known as the knot diagram. The vertices of the diagram, where the curved lines are crossed, can be considered as sites occupied by oscillators. The synchronization of these oscillators can be studied by means of a Kuramoto model. Here we propose to define some order parameters, of the complete knot diagram and of its regions, to study the synchronization of the system with regard to the different parts of it.

**Keywords**: Knots, Synchronization, Kuramoto model

**Introduction**

Networks of coupled systems have been used to model the collective dynamics of biological oscillators in self-organizing systems and in excitable media. Under certain conditions, the collective dynamics shows synchronization [1]. In fact, the importance of the synchronized behaviours in biology was already highlighted by A.T. Winfree [2] in 1967: in his paper, the author told that "the variety of biological rhythms leaves no doubt that autonomously periodic processes contribute to the coordination of life-processes."

To study the synchronized behaviour of coupled oscillators, a mathematical model, which is known as the Kuramoto model, had been developed. This model assumed the oscillators were nearly identical and that the phase of each oscillator was coupled to the collective behaviour [3,4]. Recently, some studies based on the Kuramoto model have been proposed for a modelling of neuronal synchronisation [5], after the discovery that the regions of the brain are coupled and exhibit synchronous activity [6]. As told in [5], neuronal synchronisation also plays a role in vision, movement, memory and epilepsy [7–13]. Thus, the Kuramoto model would provide a basis to modelling some phenomena of the brain.

In fact the Kuramoto model had been applied to many networks of coupled oscillators to study the collective dynamics. Studies have been performed either on regular networks, such as cubic lattices, or on random networks [3]. The small-world networks, which are intermediate of the local regular networks and the fully random networks, have been used too. The small-world networks are characterized by the fact that a high clustering, which is a characteristic of the regular networks is accompanied by a short path length, which is typically observed in random networks [14]. Phase synchronization on small-world networks emerges in the presence of even a tiny fraction of shortcuts, indicating that the same synchronizability as that of a random network can be achieved [14].

**Kuramoto oscillators and knots**

Here we propose the study of the synchronization of Kuramoto oscillators placed on the graph created by the projection of a mathematical three-dimensional knot on a bidimensional plane. This projection is known as the knot diagram. An intuitive suggestion is provided by Wikipedia [15]: think of a knot casting a shadow on the wall to have the knot diagram. Here we are talking about mathematical knots which are the embedding of a closed line in the three-dimensional space: this line can be knotted or unknotted. From a more theoretical standpoint, a knot is a homeomorphism that maps a circle into three-dimensional space and cannot be reduced to the simple circle by a continuous deformation. The difference between the mathematical and the conventional notion of a knot is that mathematical knot is closed; the conventional knot has the ends to tie or untie. The branch of mathematics that studies knots is known as knot theory [16-19].

In the upper part of Fig.1, it is shown the so-called Jordan curve of a knot, a non-self-intersecting continuous loop in the plane representing the three-dimensional knot. A "trip code" can be created considering the intersections and labelling them [17]. In the case of the diagram in Fig.1, the trip code is ABCDEFBCGEHAFGHA. The labels are letters in alphabetic order (A,B,…,H). The label is put on the vertices the first time we meet it during the trip on the knot diagram. Each letter labelling the intersections is then appearing twice in the trip code. The starting label appears three times for the convenience of the reader.
We can see from the figure that each vertex of the graph can be connected with two or four other sites, according to the curve. The knot diagram subdivides the space in several regions.

---

**Amelia Carolina Sparavigna (Correspondence)**
d002040@polito.it

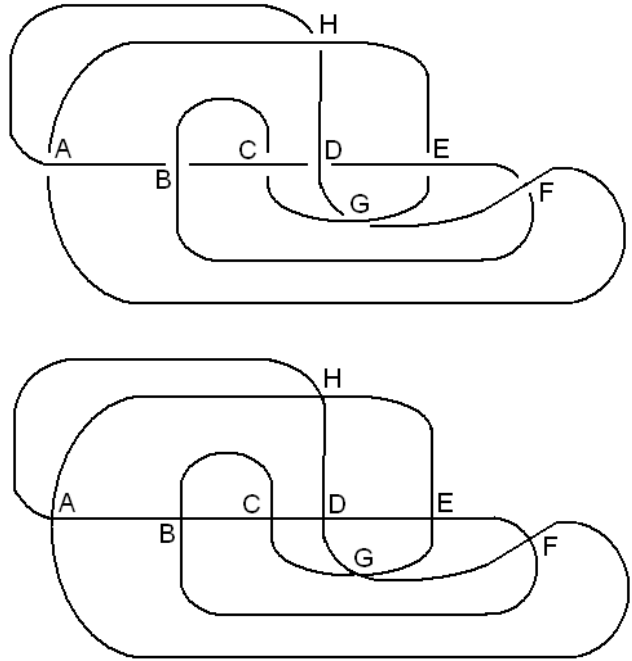

**Fig.1** – The Jordan curve of a knot and the corresponding graph. The trip code is ABCDEFBCGEHAFGHA.

Let us study the synchronization on the knot, using then a Kuramoto approach. The Kuramoto model as it stands describes oscillators of nearly natural frequencies, connected together via coupling constants. Each vertex $i$ of the knot diagram has an oscillator, the state of which is characterized by its phase $\varphi_i$. We write the set of equations of motion governing the dynamics of the $N$ oscillator system:

$$\dot{\varphi}_i + \frac{K}{2k}\sum_{j\in\Lambda_i}\sin(\varphi_i-\varphi_j) = \omega_i \qquad (1)$$

.where $\Lambda_i$ denotes the nodes connected to site $i$. For instance, let us consider site $i$=G: from the figure we see that it is connected with C, D, E and F. Then we have $\Lambda_i$=C,D,E,F, when we assume a coupling just to the nearest neighbour sites. $K$ is the coupling strength, $k$ the range of the local connection [14]. In our example $k$=4. $\omega_i$ is the intrinsic frequency of the $i$-th oscillator.

To evaluate the synchronization, an order parameter is defined [14]. Let us define it as:

$$m = \left\langle\left|\frac{1}{n}\sum_{j=1}^{n}e^{i\varphi_j}\right|\right\rangle \qquad (2)$$

.where $n$ is the number of oscillators chosen for calculation. The bracket $\langle\ \rangle$ indicates the average over time. The average over different realizations of the intrinsic frequencies, as used in [14], is not considered in the proposed calculation.

To show an example of synchronization on a knot we look at the diagram in Fig.1, having oscillators on vertices with frequencies which are ranging from 1.20 to 0.85 $s^{-1}$, reducing by steps of 0.05 $s^{-1}$, following the alphabetic order of labels ($\omega_A = 1.20\,s^{-1}$, $\omega_B = 1.15\,s^{-1}$, ..., $\omega_H = 0.85\,s^{-1}$).

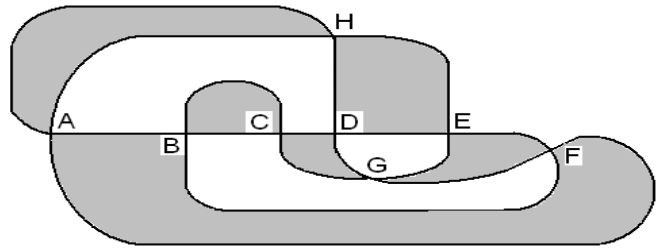

**Fig.2** – The diagram corresponding to the knot in Figure 1 divides the plane in several regions. The grey ones are here used to study the synchronization.

**Discussion and results**

We can study the role of the coupling strength *K* in the frequency synchronization. It is interesting to compare the overall behaviour of the synchronization, that we obtain using all the sites in the evaluation of the order parameter, with the synchronization of each single region of the knot diagram. As we can see in Fig.2, the curve obtained by the projection of the knot on a plane, is a curve which divides the plane is several region.

In Fig.3, the red dotted curve is showing the behaviour of the order parameter obtained from Eq.2, when it is assumed index *j* ranging from 1 (corresponding to vertex A) to *n*=8 (corresponding to vertex H). The order parameter is then evaluated considering all the vertices of the knot. The other curves are referring to the synchronization of each of the grey regions in the following way. Let us consider for instance region CDG: it is assumed in Eq.2 that *j* is ranging from 1 to 3, where 1 corresponds to C, 2 to D and 3 to G.

Let us note the different behaviour of the synchronization of regions AH and BC shown in Fig.3 (obtained from the Equation 2 of the order parameter, evaluated for *j*=A,H and *j*=B,C, respectively). As we can see from the figure, BC is synchronized for lower values of the coupling strength *K*. The reason is than BC have two vertices occupied by oscillators with the lowest possible difference in frequency, connected each other by two edges. Moreover B and C are connected by the trip code with other vertices having close frequencies. AH is a region where the vertices have oscillators with the highest frequency difference and the trip code is connecting them with oscillators quite different too.

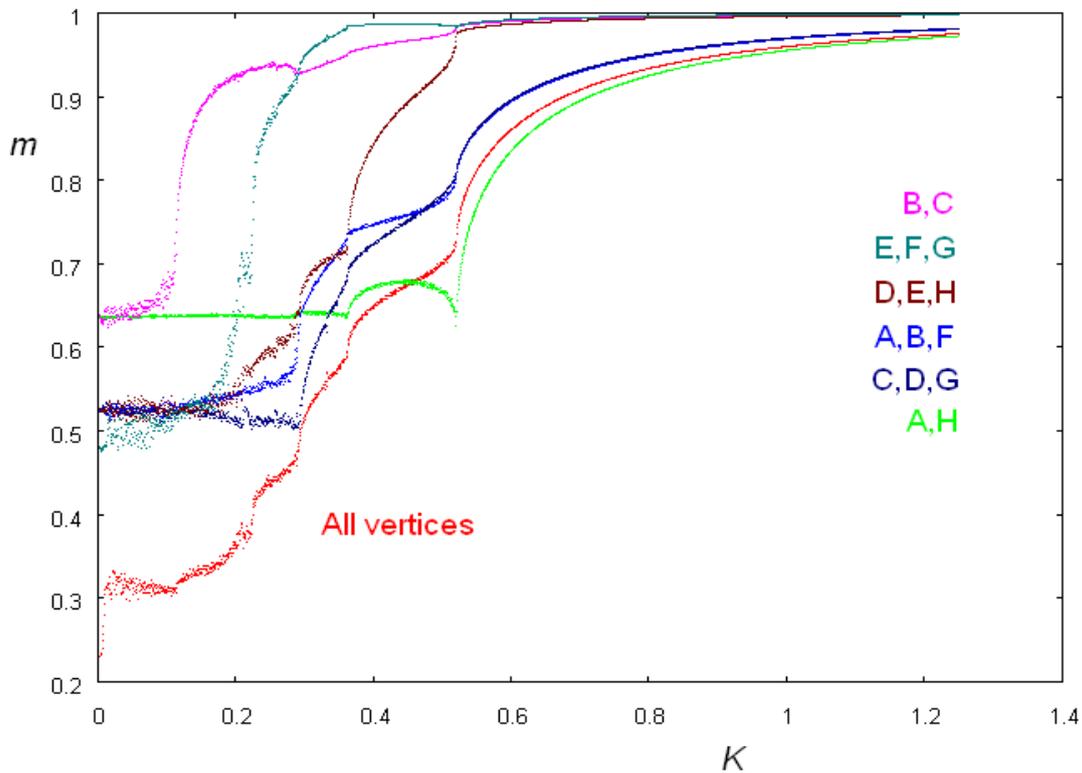

**Fig.2** – Behaviour of the order parameter defined in Equation 2 (see text for explanation) as a function of the coupling strength *K*.

The use of knot diagrams to simulate the behaviour of real systems composed by several oscillators could be useful in studying the synchronization of the system. Adjusting the frequency values according to the labels and the trip code according with the connections among sites we can obtain some order parameters, suitable to analyse the behaviour of the system. A study and analysis of existing functional integrals and invariants defined on knots is under development to verify a connection with the proposed order parameter for oscillators.